\documentclass[twocolumn,preprintnumbers,amsmath,amssymb,aps,nobibnotes,nofootinbib]{revtex4}
\usepackage{graphicx}
\usepackage{dcolumn}
\usepackage{bm}
\usepackage[a4paper, top=1.2in, bottom=0.95in, left=0.8in, right=0.8in]{geometry}

\newcommand{\dmsk}{Der\-m\'{i}\-\v{s}ek}

\newcommand{\beqn}{\begin{eqnarray}}
\newcommand{\eeqn}{\end{eqnarray}}
\def\sla#1{\setbox0=\hbox{$#1$}\dimen0=\wd0
      \setbox1=\hbox{/} \dimen1=\wd1 \ifdim\dimen0>\dimen1
      \rlap{\hbox to \dimen0{\hfil/\hfil}} #1                        \else
      \rlap{\hbox to \dimen1{\hfil$#1$\hfil}}
      /   \fi}

\newcommand{\nn}{\nonumber}

\newcommand{\al}{\alpha}
\newcommand{\be}{\beta}

\newcommand{\eps}{\epsilon}

\newcommand{\la}{\lambda}

\newcommand{\mc}{\mathcal}

\newcommand{\D}{\Delta}

\begin{document}

\preprint{TUM-HEP-681/08}
\preprint{OHSTPY-HEP-T-08-001}

\title{SUSY GUTs with Yukawa unification:\\a go/no-go study using FCNC processes}

\author{Wolfgang~Altmannshofer$^a$}
\author{Diego~Guadagnoli$^a$}%
\author{Stuart~Raby$^b$}%
\author{David~M.~Straub$^a$}%
\affiliation{%
\vspace{0.3cm}
$^a$Physik-Department, Technische Universit\"at M\"unchen,\\
D-85748 Garching, Germany\vspace{0.1cm}\\
$^b$The Ohio State University, 191 W. Woodruff Ave, Columbus, OH 43210, USA
}%

\date{\today}

\begin{abstract}
We address the viability of exact Yukawa unification in the context of general SUSY GUTs 
with universal soft-breaking sfermion and gaugino mass terms at the GUT scale. We find that 
this possibility is challenged, unless the squark spectrum is pushed well above the limits 
allowed by naturalness. This conclusion is assessed through a global fit using electroweak 
observables and quark flavour-changing neutral current (FCNC) processes. 
The problem is mostly the impossibility of accommodating simultaneously the 
bottom mass and the BR$(B \to X_s \gamma)$, after the stringent CDF upper bound on the 
decay $B_s \to \mu^+ \mu^-$ is taken into account, and under the basic assumption 
that the $b \to s \gamma$ amplitude have like sign with respect to the Standard Model one, 
as indicated by the $B \to X_s \ell^+ \ell^-$ data.

With the same strategy, we also consider the possibility of relaxing Yukawa unification 
to $b - \tau$ Yukawa unification. 
We find that with small departures from the condition $\tan \beta \simeq 50$,
holding when Yukawa unification is exact, the mentioned tension is substantially relieved.
We emphasize that in the region where fits are successful the lightest part of the SUSY 
spectrum is basically fixed by the requirements of $b - \tau$ unification and the applied 
FCNC constraints. As such, it is easily falsifiable once the LHC turns on.

\end{abstract}


\maketitle

\section{\label{sec:intro}Introduction}

The hypothesis of grand unification is able to address many of 
the unanswered questions of the Standard Model (SM), like charge quantization 
or the quantum number assignments of the SM fermions. Moreover, augmenting grand 
unified theories (GUTs) with supersymmetry (SUSY) not only stabilizes the large 
mass hierarchy between the electroweak (EW) and the GUT scale but also leads to 
the possibility of exact gauge coupling unification.

Although this unification works remarkably well in the minimal supersymmetric 
Standard Model (MSSM), in order to test the idea of grand unification one needs 
other, independent observables. A first candidate in this respect would be proton 
decay. However, the absence of a signal at proton decay experiments 
constrains mostly non-supersymmetric GUTs, whereas the strong model-dependence 
of the proton decay rate contributions from dimension-five operators 
makes it difficult to draw general conclusions
\cite{LucasRabyPRD,GotoNihei,BabuPatiWilczek,DermisekMafiRaby,Dutta04,Dutta07,Nath06,Nath07} 
on the viability of SUSY GUTs.

A different way to test SUSY GUTs is by their predictions for the masses and mixings 
of the SM fermions. In this respect, SO(10) is especially attractive because it 
unifies all quarks and leptons of one generation into a $\mathbf{16}$ representation 
of the gauge group, leading to the opportunity of a unified Yukawa coupling for 
the fermions of that generation. Though this is not phenomenologically viable for 
the two light generations, unification of the top, bottom, tau, and tau-neutrino Yukawa 
couplings might be possible if $\tan\beta$, the ratio of the vacuum expectation 
values of the two Higgs doublets, is close to 50.

It must be taken into account, however, that the success of this Yukawa unification sensitively 
depends not only on $\tan \beta$, but also on the SUSY spectrum and parameters, 
because the Yukawa couplings are much more sensitive to weak scale threshold corrections 
than are the gauge couplings. In the absence of a clear signal in favour of supersymmetry 
or an {\em a priori} knowledge of the SUSY spectrum, one would need a set of
additional observables to sufficiently constrain the allowed ranges for the 
SUSY spectrum itself, in order to test GUT predictions for fermion masses. It turns 
out that flavour-changing neutral current (FCNC) processes --~loop-suppressed observables 
that are highly sensitive to SUSY particle contributions~-- are especially suited 
for this purpose.

In \cite{AABuGuS}, an SO(10) SUSY GUT model proposed by \dmsk\ and Raby (DR) \cite{DR05}, 
and featuring Yukawa unification, has been reconsidered in a global analysis in light 
of all the most precise data on FCNCs in the quark sector. While the model successfully 
describes EW observables as well as quark and lepton masses and 
mixings \cite{DR05,DR06}, in \cite{AABuGuS} it was found that the simultaneous 
description of these observables {\em and} all the FCNC processes considered is 
impossible unless the squark masses are pushed well above the limits allowed by 
naturalness and within reach of the Large Hadron Collider (LHC).

The aim of this Letter is twofold. First, we show that the problem pointed out 
in \cite{AABuGuS} and mentioned above is a general feature of SUSY GUT models 
with Yukawa unification and universal sfermion and gaugino mass terms at the GUT scale, 
thus challenging the viability of these hypotheses, when considered together.
Our second aim is then to explore a possible remedy, namely relaxing the hypothesis 
of Yukawa unification in favour of the less restrictive $t - \nu$ and $b - \tau$ Yukawa 
unifications. The departure from exact Yukawa unification can be quantified by the 
parametric departure from the condition $\tan \beta \simeq 50$. As clarified below, this 
case will be relevant not only for SU(5), but also for SO(10). 
This study will allow us to address 
the question whether a range of large $\tan \beta \lesssim 50$ exists, where a successful 
prediction for the bottom mass {\em and} full compatibility with quark FCNCs are possible 
at the same time.

\section{\label{sec:yukawa}Yukawa unification and FCNCs}

It is well-known \cite{BDR1,BDR2} that, under the assumptions of a universal sfermion mass $m_{16}$ 
and a universal gaugino mass $m_{1/2}$ at the GUT scale, and with a positive $\mu$ 
parameter, Yukawa unification prefers the region in MSSM parameter space characterized 
by the relations
\beqn
-A_0\approx 2\,m_{16}, ~~ \mu,m_{1/2} \ll m_{16},
\label{ISMH}
\eeqn
because they ensure a cancellation of potentially large $\tan \beta$-enhanced 
SUSY threshold corrections to the bottom quark mass \cite{HRS}, which could otherwise 
spoil the Yukawa unification. Through renormalization group effects, these relations 
lead to an inverted scalar mass hierarchy (ISMH) \cite{BFPZ}, i.e., light third generation 
and heavy first and second generation sfermions.

Relations (\ref{ISMH}), together with the large value of $\tan \beta \approx 50$ 
required for Yukawa unification, have an important impact on the SUSY spectrum and 
on the predictions for FCNCs. In particular, the branching ratio of the decay 
$B_s \to \mu^+ \mu^-$ receives large $\tan \beta$-enhanced contributions from Higgs-mediated 
neutral currents that are proportional to $A_t^2 (\tan \beta)^6/M_A^4$ \cite{BabuKolda,Gaur}. 
With large $\tan \beta$ and a large trilinear coupling $A_t$ following from 
relations (\ref{ISMH}), the stringent most recent experimental bound \cite{Bsmumu-CDF}
\beqn
\text{BR}(B_s\to\mu^+\mu^-)_\text{exp} < 5.8 \times 10^{-8} ~~\text{(95\% C.L.})
\eeqn
can only be met with quite heavy $A^0$, $H^0$, and $H^+$ Higgs bosons.

Another important process in this respect is the tree-level decay $B^+ \to \tau^+ \nu$. 
Using the SM fit value for the CKM matrix element $V_{ub}$ \cite{UTfit,CKMfitter} 
one obtains a SM prediction for the branching ratio\footnote{This prediction is obtained by 
normalizing the branching ratio to $\D M_d$ \cite{Ikado}. The value in eq. (\ref{BtaunuSM})
agrees well with those reported in \cite{AABuGuS,CMW,UTfit}.}
\beqn
\text{BR}(B^+ \to \tau^+ \nu)_\text{SM} = (0.82 \pm 0.11)\times 10^{-4}\phantom{~.}
\label{BtaunuSM}
\eeqn
that is quite low compared to the experimental value \cite{HFAG,Babar-Btaunu,Belle-Btaunu}
\beqn
\text{BR}(B^+ \to \tau^+ \nu)_\text{exp} = (1.41 \pm 0.43)\times 10^{-4}~.
\label{BtaunuExp}
\eeqn
In the MSSM with large $\tan \beta$, the dominant additional contribution to this process 
comes from charged Higgs bosons and is found to interfere always destructively \cite{Hou} with 
the SM contribution, thus further reducing the theory prediction. 
Hence, similarly to $B_s \to \mu^+ \mu^-$, the $B^+\to\tau^+\nu$ decay requires 
a heavy Higgs spectrum to be in agreement with the experimental data. However, given the 
large experimental uncertainty in (\ref{BtaunuExp}), the $B_s \to \mu^+ \mu^-$ constraint 
turns out usually to be more stringent.

Finally, a very important constraint is the inclusive decay $B \to X_s \gamma$, which 
receives the dominant SUSY contributions from a chargino--stop loop and a top--charged Higgs 
loop.\footnote{In the actual numerical analysis we include all the relevant contributions, 
in particular gluino-down squark loops as well. The latter are found to play a negligible role.} 
The chargino contribution is $\tan \beta$-enhanced and, with the large negative 
trilinear parameters implied by relations (\ref{ISMH}), adds destructively to the 
SM branching ratio. The charged Higgs contribution, on the other hand, adds constructively 
to the branching ratio, but is suppressed by the heavy Higgs masses required to be consistent 
with $B_s \to \mu^+ \mu^-$. Thus a near cancellation between the two contributions, which is 
necessary in view of the good agreement between the experimental determination \cite{Barberio07}
\beqn
\label{bsgamma-exp}
&&\text{BR}(B \to X_s \gamma)_\text{exp} \nn \\
&&\hspace{1cm}= (3.55 \pm 0.24 ^{+0.09}_{-0.10} \pm 0.03) \times 10^{-4}~~~~
\eeqn
and the SM prediction \cite{Misiak-NNLO}
\beqn
\text{BR}(B \to X_s \gamma)_\text{SM} = (3.15 \pm 0.23) \times 10^{-4}~,
\label{bsgamma-SM}
\eeqn 
is difficult to achieve. 

Note that the solution with the chargino contribution so large that the sign 
of the $b \to s \gamma$ amplitude is reversed \cite{BlazekRaby-bsgamma} is challenged in our framework 
by the experimental data on $\text{BR}(B \to X_s \ell^+ \ell^-)$ \cite{GHM,ALGH,LPV}. In fact, it 
would lead to a 3$\sigma$ discrepancy between the prediction and the experimental figure for this 
branching ratio.\footnote{This statement holds, barring non-negligible new physics 
contributions to the Wilson coefficients $\tilde{C}^{\rm eff}_{9,10}$ (see Ref. \cite{GHM}), 
which is impossible in our case.}

Leaving aside, for the moment, the possibility of a sign flip in the $b \to s \gamma$ amplitude 
(we will return to this issue in section \ref{sec:proc}), the above discussion implies that a 
tension between the prediction for $B \to X_s \gamma$ and the bound on $B_s \to \mu^+ \mu^-$ 
should generally be expected in models with Yukawa unification, as a direct consequence of 
relations (\ref{ISMH}) and the large value of $\tan \beta$. By the nature of the argument, this 
tension should be completely independent of the mechanism (flavour symmetries or other) embedded 
in the SUSY GUT to explain the light quark masses and mixings. In section~\ref{sec:results} 
we will come back to this issue, showing that indeed this tension occurs generally in SUSY GUT 
models with Yukawa unification and quantifying the tension numerically with a $\chi^2$-procedure.

Technically, the most immediate potential remedy to the above mentioned problem seems to be 
to lower $\tan \beta$. This in fact alleviates the pressure from $B_s \to \mu^+ \mu^-$, 
permitting in turn larger Higgs and smaller chargino contributions to $B \to X_s \gamma$ 
and thereby making possible that those two contributions indeed cancel to a large extent.

Lowering $\tan \beta$ means breaking the unification of the top and bottom 
Yukawa couplings, so that the full Yukawa unification is relaxed to $b - \tau$ unification, 
occurring, e.g., in SU(5). Such breaking of $t - b$ unification is actually also a general 
possibility in SO(10) SUSY GUTs once all 
the representations needed for a realistic GUT-breaking sector are taken into account. For 
example, the ``minimal breaking scheme'' introduced by Barr and Raby \cite{BarrRaby} requires 
the presence of a $\mathbf{16}'_H$ spinor. In this framework, the MSSM Higgs doublet $H_d$ 
can naturally be a mixture between a doublet contained in the same $\mathbf{10}_H$ representation 
as the doublet $H_u$ and one doublet contained in this $\mathbf{16}'_H$ spinor, since 
the two have the same quantum numbers. One then has \cite{BabuPatiWilczek,AlbrightBarr,DermisekMafiRaby}
\beqn
H_u &=& H(\mathbf{10}_H)~,\nonumber \\
H_d &=& \overline H(\mathbf{10}_H) \cos \gamma + \overline H(\mathbf{16}'_H) \sin \gamma~.
\label{HuHd}
\eeqn
Consequently, the Yukawa unification relation $\la_t = \la_b$ is effectively broken to 
\beqn
\frac{\la_b}{\la_t} = \cos \gamma~.
\label{lad/lau}
\eeqn
At the EW scale, this relation leads to a value of $\tan\beta \lesssim 50$ 
parametrically smaller than in the exact unification case, depending on the amount of mixing 
in the second of eqs. (\ref{HuHd}).

We would like to emphasize that the two cases of SU(5) and SO(10) with minimal 
breaking scheme mentioned above are just intended as examples. Our analysis will be completely 
general in SUSY GUTs with $b - \tau$ unification.

It should be stressed as well that, even without $t - b$ unification, SUSY GUT models with 
$b - \tau$ unification maintain in fact most of their predictivity, since the relation between 
the $b$ and $\tau$ Yukawa couplings {\em requires} the ISMH relations, eq. (\ref{ISMH}), to be 
satisfied in order to obtain a correct prediction for $m_b$. 
In addition, a crucial observation is that $b - \tau$ unification requires $\tan \beta$ either 
close to unity (which is however excluded by the Higgs mass bound \cite{LEPHiggs}) or O(50), 
because otherwise the predicted bottom quark mass is in general too large \cite{CPW,ananthanarayan}.
Although the case $\tan \beta =$ O(50) can be significantly modified by the $\tan \beta$-enhanced 
threshold corrections to $m_b$ mentioned above, $b - \tau$ unification is difficult to achieve for 
$\tan\beta \lesssim 35$. Therefore the strategy to lower $\tan \beta$ is not a trivial one 
in our context, since $b - \tau$ unification pushes by itself $\tan \beta$ to high values.

With the above arguments, departure from third generation Yukawa unification and restriction 
to $b - \tau$ unification seems to be a promising approach to retain the 
predictivity of GUT models, while at the same time possibly removing tensions in FCNC observables, 
thanks to $\tan \beta < 50$. The rest of our Letter is thus an attempt to address the following 
two questions:
\begin{itemize}
\item Is the tension between FCNC observables a general feature of GUT models 
with third generation Yukawa unification and universal masses for sfermions and gauginos 
at the GUT scale;
\item Is this tension relieved when $\tan \beta$ is (slightly) below 50, 
i.e., if one moves from exact Yukawa unification but retains $b - \tau$ unification.
\end{itemize}
These issues will be studied through a numerical procedure to be described below.

\section{\label{sec:proc}Procedure}

We assume, at scales higher than the GUT scale $M_G$, the existence of a grand unified 
group entailing $b - \tau$ unification. Beneath $M_G$ the grand unified group 
is broken\footnote{For our purposes, this can be assumed to happen in one single step.} 
to the SM group $G_{\rm SM} \equiv SU(3)_c \times SU(2)_L \times U(1)_Y$ and the MSSM 
running is performed down to the EW scale. As for the GUT scale boundary conditions to 
this running, we include a unified gauge coupling $\alpha_G$, allowing for a GUT scale 
threshold correction $\eps_3$ to the strong coupling constant, the Yukawa couplings for 
up- and down-type fermions of the third 
generation $\la_t$ and $\la_b$,\footnote{On the Yukawa couplings of the lightest two 
generations we will comment later on in this section.} and a right-handed neutrino mass 
$M_R$. At $M_G$ we also assume a soft SUSY-breaking sector, consisting of a 
universal trilinear coupling $A_0$, a universal sfermion mass $m_{16}$, a universal gaugino 
mass $m_{1/2}$, as well as non-universal Higgs mass parameters $m_{H_u}$, $m_{H_d}$.

We run all the parameters using one-loop RGEs for the soft sector and two-loop RGEs for the 
Yukawa and gauge couplings \cite{MartinVaughn}.
To take correctly into account the effects of right-handed neutrinos present in SO(10) and 
required for the see-saw mechanism, we include the contribution of  a third-generation neutrino 
Yukawa coupling (with initial condition $\lambda_{\nu_\tau} = \lambda_t$) in all RGEs between 
$M_G$ and $M_R$ \cite{Hisano95,AKLR02,Petcov03}. 
In our framework, there are thus no potentially large logarithmic GUT scale threshold 
corrections to either Yukawa unification or to Higgs splitting, which would be present if 
such contribution were neglected in the RGEs.\footnote{In contrast with statements 
made in the literature and in accord with the results of \cite{DR05}, we find that neutrino 
Yukawa effects are not sufficient to explain the large Higgs splitting required for successful 
EWSB to occur (see \cite{DR05}, footnote 15).} The remaining GUT scale threshold corrections 
to the Yukawa couplings are expected to be small \cite{BDR2}. 

At the EW scale, we finally have the two additional free parameters $\tan \beta$ and $\mu$. 
The total number of free parameters is then 13 and they are collected in table~\ref{tab:parameters}.
\begin{table}[ht]
\begin{tabular}{|lcc|}
\hline
Sector & ~~\#~~ & Parameters \\
\hline \hline
gauge & 3 & $\alpha_G$, $M_G$, $\eps_3$ \\
SUSY & 5 & $m_{16}$, $m_{1/2}$, $A_0$, $m_{H_u}$, $m_{H_d}$ \\
Yukawas & 2 & $\la_t$, $\la_b$ \\
neutrino & 1 & $M_{R}$ \\
SUSY (EW scale) & 2 & $\tan \beta$, $\mu$ \\
\hline
\end{tabular}
\caption{Model parameters. Unless explicitly stated, they are intended at the GUT scale.}
\label{tab:parameters}
\end{table}

After calculating the SUSY and Higgs spectra\footnote{We calculate the Higgs VEVs and $M_A$ 
following \cite{PBMZ} and use {\tt FeynHiggs} \cite{FeynHiggs1,FeynHiggs2,FeynHiggs3,FeynHiggs4} 
to obtain the light Higgs mass.} and the threshold corrections to third generation fermion 
masses \cite{PBMZ}, we evaluate the flavour-changing observables using the effective Lagrangian 
approach of \cite{BCRSbig}. Thereafter, in order to have a quantitative 
test of the model, we construct a $\chi^2$ function defined as
\beqn
\chi^2[\vec\vartheta] \equiv \sum_{i = 1}^{N_{\rm obs}}
\frac{(f_i[\vec\vartheta] -\mc O_i)^2}
{(\sigma_i^2)_{\rm exp} + (\sigma_i^2)_{\rm theo}}~,
\label{chi2}
\eeqn
composed of the quantities given in tables \ref{tab:obs-EW} and \ref{tab:obs-FC}.
\begin{table}[ht]
\begin{tabular}{|lc||lc|}
\hline
Observable & ~Value($\sigma_{\rm exp}$)~ & Observable & ~Lower Bound~ \\
\hline \hline
$M_W$ & $80.403(29)$ & $M_{h_0}$ & $114.4$ \\
$M_Z$ & $91.1876(21)$ & $M_{\tilde \chi^+}$ & $104$ \\
$10^{5} G_\mu$ & $1.16637(1)$ & $M_{\tilde t}$ & $95.7$ \\
$1/\al_\text{em}$ & $137.036(0)$ & & \\
$\al_s(M_Z)$ & $0.1176(20)$ & & \\
$M_t$ & $170.9(1.8)$ & & \\
$m_b(m_b)$ & $4.20(7)$ & & \\
$M_\tau$ & $1.777(0)$ & & \\
\hline
\end{tabular}
\caption{Flavour conserving observables \cite{PDBook,CDFD0mt} used in the fit. 
Dimensionful quantities are expressed in powers of GeV.}
\label{tab:obs-EW}
\end{table}
\begin{table}[t]
\begin{tabular}{|lcr|}
\hline
Observable & ~~Value($\sigma_{\rm exp}$)($\sigma_{\rm theo}$)~~ & Ref. \\
\hline \hline
$\D M_s / \D M_d$ & 35.1(0.4)(3.6) & \cite{Barberio07,CDF-DMs} \\
$10^4$ BR$(B \to X_s \gamma)$ & 3.55(26)(46) & \cite{Barberio07} \\
$10^6$ BR$(B \to X_s \ell^+ \ell^-)$ & 1.60(51)(40) & \cite{Babar-bsll,Belle-bsll} \\
$10^4$ BR$(B^+ \to \tau^+ \nu)$ & 1.41(43)(26) & \cite{HFAG} \\
BR$(B_s \to \mu^+ \mu^-)$ & $< 5.8 \times 10^{-8}$ & \cite{Bsmumu-CDF} \\
\hline
\end{tabular}
\caption{Flavour-changing observables used in the fit. The BR$(B \to X_s \ell^+ \ell^-)$ 
is intended in the range $q^2_{\ell^+ \ell^-} \in [1,6]$ GeV$^2$.}
\label{tab:obs-FC}
\end{table}

\noindent In eq. (\ref{chi2}) $\mc O_i$ indicates the experimental value of the 
observables listed in tables \ref{tab:obs-EW} and \ref{tab:obs-FC} and 
$f_i[\vec \vartheta]$ the corresponding theoretical prediction, which will be function 
of the model parameters listed in table \ref{tab:parameters}, collectively indicated 
with $\vec \vartheta$. The $\chi^2$ function is minimized upon variation of the model 
parameters, using the minimization algorithm {\tt MIGRAD}, which is part of the {\tt CERNlib} 
library \cite{CERNlib}.

Some comments are in order on the determination of the errors.
First, one can note that among the observables in table \ref{tab:obs-EW}, some have a 
negligible experimental error. In this case, we took as overall uncertainly 0.5\% of the 
experimental value, which we consider a realistic estimate of the numerical error 
associated with the calculations. Concerning the theoretical errors on the flavour 
observables (table \ref{tab:obs-FC}), we note the following: the error on 
$\D M_s / \D M_d$ takes into account the SM contribution, dominated by $\xi^2$ 
and the new physics contributions, dominated by scalar operators;
the error on BR$(B^+ \to \tau^+ \nu)$, after normalization by $\D M_d$ \cite{Ikado}, 
is dominated by the lattice ``bag'' parameter $\hat B_d$ and the relevant CKM 
entries; the error on BR$(B \to X_s \gamma)$ is taken as twice\footnote{\label{foot}
This choice is quite conservative, considering that in our case variations of 
the calculated BR$(B \to X_s \gamma)$ upon variation of the SUSY matching scale in the 
huge range [0.1, 1] TeV are typically around 4\%. However, we feel it is justified
in the case of large cancellations among new physics contributions.} 
the total theoretical error associated with the SM calculation \cite{Misiak-NNLO}; 
finally the error on BR$(B \to X_s \ell^+ \ell^-)$ is taken as 25\% of the experimental 
result, and is estimated from the spread of the theoretical predictions after 
variations of the scale of matching of the SUSY contributions.

In evaluating the $\chi^2$ function, we also included the bounds reported in tables 
\ref{tab:obs-EW} and \ref{tab:obs-FC}. These constraints are in the form of suitably 
smoothened step functions, which are added to the $\chi^2$-function of eq. (\ref{chi2}). 
If any of the constraints is violated, the step functions add a large positive number 
to the $\chi^2$, while for respected constraints the returned value is zero, so that 
the $\chi^2$ is set back to its `unbiased' definition (\ref{chi2}).

A step function was also included in order to enforce the desired sign for the 
$b \to s \gamma$ calculated amplitude, thus permitting to systematically explore both 
cases of like sign or flipped sign with respect to the SM one.
In the case of flipped sign, large SUSY contributions are necessary such that SUSY 
is not quite a correction to the SM result, but rather the opposite.
As a consequence, one would need a theoretical control on the SUSY part at least as good 
as that on the pure SM calculation. In the absence of this knowledge, the amplitude in the 
flipped-sign case is generally very sensitive to variations of the matching scale, and the 
associated theoretical error hard to control. In order to be able to estimate 
as reliably as possible the $b \to s \gamma$ amplitude also in the flipped-sign case, we have 
taken advantage of the {\tt SusyBSG} code \cite{SusyBSG}, which is directly called by the 
fitting procedure.

As already mentioned in section \ref{sec:yukawa}, the scenario with flipped $b \to s \gamma$ 
amplitude leads to $\chi^2 \gtrsim 9$ solely on account of the implied 3$\sigma$ 
discrepancy in $B \to X_s \ell^+ \ell^-$. We calculated the BR$(B \to X_s \ell^+ \ell^-)$
using the results of Ref. \cite{Huber-et-al}. We will address the scenario with flipped 
$b \to s \gamma$ amplitude quantitatively in section \ref{sec:results}.

\bigskip

An important observation is in order at this point, justifying why our analysis should 
be valid for {\em any} SUSY GUT model with $b - \tau$ unification and universal soft
terms as in table \ref{tab:parameters}. As already mentioned in 
section \ref{sec:yukawa}, any such model prefers the region of parameter space leading 
to ISMH, i.e., third generation sfermion masses much lighter than first and second 
generation ones, the latter being of O($m_{16}$). One should as well consider that, 
assuming hierarchical Yukawa matrices and large $\tan \beta$, it is sufficient to 
include only the 33-elements of Yukawa matrices in the RGEs, that is, take GUT-scale boundary 
conditions for the Yukawa couplings as
\beqn
Y_{u,d} = {\rm diag}(0,~0,~\la_{t,b})~.
\label{yud-approx}
\eeqn
This will have a negligible effect on the determination of the first and second generation 
sfermion masses, given their heaviness.
This observation makes it possible to separate the effects of specific Yukawa textures, 
which depend on the flavour model one embeds into the SUSY GUT, from those genuinely 
due to the unification of Yukawa couplings. The adoption of this strategy brings us to 
the following remarks:
\begin{itemize}

\item Given the approximation we adopt for the initial conditions on the Yukawas, we do 
not need to assume any particular flavour model. The low-energy input of the CKM matrix 
and of the fermion masses other than third generation ones, necessary for the calculation 
of many among the observables included in the fit, is then taken directly from 
experiment\footnote{Specifically, the CKM input is taken from the new physics independent 
CKM fit \cite{UTfit}.}.

\item  Due to ISMH, the lighter stop is always the lightest sfermion and in fact 
its tree-level mass can be very small. Therefore, we include the one-loop corrections to the 
light stop mass to ensure our solutions are consistent with the lower bound in 
table \ref{tab:obs-EW}. In practice, due to its lightness, we calculate the stop pole mass 
with the same accuracy as the pole masses of the third generation fermions, the $W$, $Z$, 
and the Higgs bosons.

\item The validity of our approximation, eq. (\ref{yud-approx}), was checked by performing 
the full analysis also with a SUSY GUT model with specific flavour textures, namely the 
DR model \cite{DR05} (with Yukawa unification relaxed as in eq. (\ref{lad/lau})). Our results 
were not significantly affected.

\end{itemize}

\section{\label{sec:results}Results}

\begin{figure*}[htb]
\begin{center}
\includegraphics[width=0.98 \textwidth]{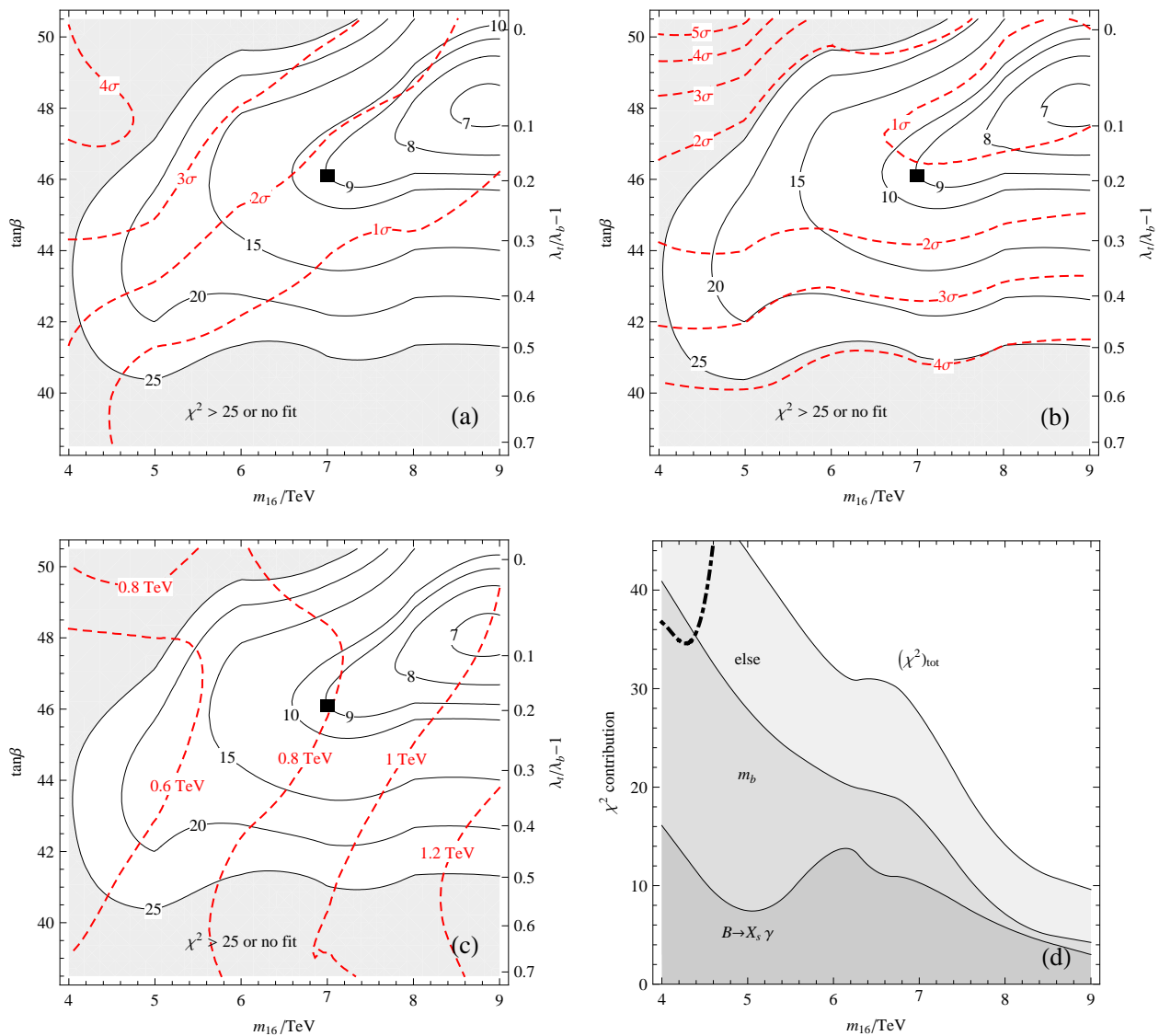}
\caption{Panels (a)-(c): $\chi^2$ contours (solid lines) in the $m_{16}$ vs. $\tan \beta$ plane.
Superimposed as dashed lines are the pulls for BR$(B \to X_s \gamma)$ (panel (a)) and for $m_b$ 
(panel (b)) and the lightest stop mass contours (panel (c)).
Panel (d): $\chi^2$ contributions vs. $m_{16}$ in the special case of exact Yukawa unification.
All the plots assume a SM-like sign for the $b \to s \gamma$ amplitude, except for panel (d), 
where also the total $\chi^2$ for the flipped-sign case is shown as a dot-dashed line.}
\label{fig}
\end{center}
\end{figure*}
In order to address the questions outlined in the introduction, we have explored in the 
$m_{16}$ vs. $\la_t/\la_b$ plane the class of GUT models defined in section \ref{sec:proc}.
Concretely, we fixed $m_{16}$ to values $\ge 4$ TeV and $\la_t/\la_b$ to values $\ge 1$ 
(roughly equivalent to fixing $\tan \beta$ to values $\le 50$) and minimized the $\chi^2$ 
function (\ref{chi2}) upon variation of the remaining model 
parameters. The minimum $\chi^2$ value provides then a quantitative test of the 
performance of the model.\footnote{Of course such test cannot be attached a statistically 
rigorous meaning, since, e.g., the $\chi^2$-entries are not all independently measured 
observables.} 
The results of our survey are reported in 
the four panels of Fig. \ref{fig}. In particular, panels (a) to (c) report the $\chi^2$ contours 
as solid lines in the $m_{16}$ vs. $\tan \beta$ plane. As reference, also the values of 
$\la_t/\la_b$ are reported on a right-hand vertical scale. Superimposed to the 
$\chi^2$ contours are: in panels (a) and (b), the deviations of respectively 
BR$(B \to X_s \gamma)$ and $m_b$ from the central values in 
tables \ref{tab:obs-EW}-\ref{tab:obs-FC} in units of the total error; in panel (c) the 
mass contours of the lightest stop. Finally, panel (d) shows the $\chi^2$ contributions from 
BR$(B \to X_s \gamma)$, $m_b$ and all the rest (as three stacked contributions, represented 
by solid lines) vs. $m_{16}$ in the special case of $\la_t/\la_b = 1$, corresponding to 
exact Yukawa unification.
\begin{table*}[t]
\begin{tabular}{|lccc|}
\hline
Observable  &  Exp.  &  Fit  &  Pull  \\
\hline\hline
$M_W$  &  80.403  &  80.56  &  0.4  \\
$M_Z$  &  91.1876  &  90.73  &  \textbf{1.0}  \\
$10^{5}\; G_\mu$  &  1.16637  &  1.164  &  0.3  \\
$1/\alpha_\text{em}$  &  137.036  &  136.5  &  0.8  \\
$\alpha_s(M_Z)$  &  0.1176  &  0.1159  &  0.8  \\
$M_t$  &  170.9  &  171.3  &  0.2  \\
$m_b(m_b)$  &  4.20  &  4.28  &  \textbf{1.1}  \\
$M_\tau$  &  1.777  &  1.77  &  0.4  \\
$10^{4}\; \text{BR} (B \to X_s \gamma)$  &  3.55  &  2.72  &  \textbf{1.6}  \\
$10^{6}\; \text{BR} (B \to X_s \ell^+\ell^-)$  &  1.60  &  1.62  &  0.0  \\
$\Delta M_s / \Delta M_d$  &  35.05  &  32.4  &  0.7  \\
$10^{4}\; \text{BR} (B^+ \to \tau^+\nu)$  &  1.41  &  0.726  &  \textbf{1.4}  \\
$10^{8}\; \text{BR} (B_s \to \mu^+\mu^-)$  &  $<5.8 $ &  3.35  &  --  \\
\hline
\multicolumn{3}{|r}{total $\chi^2$:}  &  \textbf{8.78} \\
\hline
\end{tabular}
\hspace{2cm}
\begin{tabular}{|lc|lc|}
\hline
\multicolumn{2}{|l}{Input parameters} & \multicolumn{2}{|l|}{Mass predictions} \\
\hline\hline
$m_{16}$  &  $7000$  &  $M_{h^0}$  &  121.5  \\
$\mu$  &  $1369$  &  $M_{H^0}$  &  585  \\
$M_{1/2}$  &  $143$  &  $M_{A}$  &  586  \\
$A_0$  &  $-14301$  &  $M_{H^+}$  &  599  \\
$\tan\beta$  &  $46.1$  &  $m_{\tilde t_1}$  &  783  \\
$1/\alpha_G$  &  $24.7$  &  $m_{\tilde t_2}$  &  1728  \\
$M_G / 10^{16}$  &  $3.67$  &  $m_{\tilde b_1}$  &  1695  \\
$\epsilon_3 / \%$  &  $-4.91$  &  $m_{\tilde b_2}$  &  2378  \\
$(m_{H_u}/m_{16})^2$  &  $1.616$  &  $m_{\tilde \tau_1}$  &  3297  \\
$(m_{H_d}/m_{16})^2$  &  $1.638$  &  $m_{\tilde\chi^0_1}$  &  58.8  \\
$M_{R} / 10^{13}$  &  $8.27$  &  $m_{\tilde\chi^0_2}$  &  117.0  \\
$\lambda_u$  &  $0.608$  &  $m_{\tilde\chi^+_1}$  &  117.0  \\
$\lambda_d$  &  $0.515$  &  $M_{\tilde g}$  &  470  \\
\hline
\end{tabular}
\caption{Example of successful fit in the region with $b -\tau$ unification. Dimensionful 
quantities are expressed in powers of GeV. Higgs, lightest stop and gluino masses are pole 
masses, while the rest are running masses evaluated at $M_Z$.}
\label{tab:fit-example}
\end{table*}
\begin{table*}[t]
\begin{tabular}{|lccc|}
\hline
Observable  &  Exp.  &  Fit  &  Pull  \\
\hline\hline
$M_W$  &  80.403  &  80.32  &  0.2  \\
$M_Z$  &  91.1876  &  90.63  &  \textbf{1.2}  \\
$10^{5}\; G_\mu$  &  1.16637  &  1.162  &  0.7  \\
$1/\alpha_\text{em}$  &  137.036  &  136.4  &  \textbf{1.0}  \\
$\alpha_s(M_Z)$  &  0.1176  &  0.1144  &  \textbf{1.5}  \\
$M_t$  &  170.9  &  171.5  &  0.3  \\
$m_b(m_b)$  &  4.2  &  4.41  &  \textbf{2.9}  \\
$M_\tau$  &  1.777  &  1.77  &  0.7  \\
$10^{4}\; \text{BR} (B \to X_s \gamma)$  &  3.55  &  3.69  &  0.3  \\
$10^{6}\; \text{BR} (B \to X_s \ell^+\ell^-)$  &  1.60  &  4.41  &  \textbf{4.3}  \\
$\Delta M_s / \Delta M_d$  &  35.05  &  32.5  &  0.7  \\
$10^{4}\; \text{BR} (B^+ \to \tau^+\nu)$  &  1.41  &  0.561  &  \textbf{1.7}  \\
$10^{8}\; \text{BR} (B_s \to \mu^+\mu^-)$  &  $<5.8 $ &  5.0  &  --  \\
\hline
\multicolumn{3}{|r}{total $\chi^2$:}  &  \textbf{36.5} \\
\hline
\end{tabular}
\hspace{2cm}
\begin{tabular}{|lc|lc|}
\hline
\multicolumn{2}{|l}{Input parameters} & \multicolumn{2}{|l|}{Mass predictions} \\
\hline\hline
$m_{16}$  &  $4000$  &  $M_{h^0}$  &  114.5  \\
$\mu$  &  $249$  &  $M_{H^0}$  &  509  \\
$M_{1/2}$  &  $149$  &  $M_{A}$  &  510  \\
$A_0$  &  $-7989$  &  $M_{H^+}$  &  519  \\
$\tan\beta$  &  $50.2$  &  $m_{\tilde t_1}$  &  425  \\
$1/\alpha_G$  &  $24.7$  &  $m_{\tilde t_2}$  &  823  \\
$M_G / 10^{16}$  &  $2.24$  &  $m_{\tilde b_1}$  &  680  \\
$\epsilon_3 / \%$  &  $-3.78$  &  $m_{\tilde b_2}$  &  759  \\
$(m_{H_u}/m_{16})^2$  &  $1.643$  &  $m_{\tilde \tau_1}$  &  1402  \\
$(m_{H_d}/m_{16})^2$  &  $1.908$  &  $m_{\tilde\chi^0_1}$  &  59.3  \\
$M_{R} / 10^{13}$  &  $3.30$  &  $m_{\tilde\chi^0_2}$  &  109.3  \\
$\lambda_u$  &  $0.643$  &  $m_{\tilde\chi^+_1}$  &  108.9  \\
$\lambda_d$  &  $0.643$  &  $M_{\tilde g}$  &  461  \\
\hline
\end{tabular}
\caption{Representative fit obtained in the region with $t - b -\tau$ unification and flipped $b \to s \gamma$ amplitude. 
Conventions as in table \ref{tab:fit-example}.}
\label{tab:flipped-fit-example}
\end{table*}

Various comments are in order on these plots.
\begin{enumerate}

\item Panel (d) shows that, for any $m_{16} \lesssim 9$ TeV, the $\chi^2$ contribution from 
$B \to X_s \gamma$ alone is no less than roughly 4, corresponding to no less than $2\sigma$ deviation 
from the result of eq. (\ref{bsgamma-exp}). Therefore, in the case of Yukawa unification, agreement among
FCNCs is only achieved at the price of decoupling in the scalar 
sector.\footnote{For similar findings in the context of Bayesian analyses of the CMSSM, see, e.g.,
Ref. \cite{Roszkowski-CMSSM}.}
One should note in this respect the quite conservative choice of the $B \to X_s \gamma$ error, already 
mentioned in footnote \ref{foot}. The apparent non-monotonic behaviour of the $B \to X_s \gamma$ 
$\chi^2$-profile is due to the fact that, for $m_{16} \lesssim 6$ TeV, the model prediction becomes 
so bad that the minimization algorithm prefers to sacrifice the prediction for $m_b$ (which in fact 
gets much worse).

\item For $m_{16} \lesssim 4.7$ TeV, fits usually prefer the flipped-sign solution for the 
$b \to s \gamma$ amplitude, discussed in section \ref{sec:proc}. However, at the quantitative level, 
this requirement (implying a quite light stop mass) turns out to be difficult to reconcile with that 
of successful predictions for the bottom mass and/or for EW observables. The solutions we found have 
$\chi^2 \gtrsim 14$ as result of the contributions from EW observables and $m_b$ alone. On top of it 
one has to add the $\chi^2$ contributions from FCNCs, being $\gtrsim 9$ solely on account of 
the $3\sigma$ discrepancy in BR$(B \to X_s \ell^+\ell^-)$. The total $\chi^2$ for the 
fits with flipped-sign $b \to s \gamma$ amplitude is reported in figure \ref{fig} (d) as a dot-dashed line.

\item \label{itemI}From panels (a)-(c) one can note a region of successful fits for $m_{16} \gtrsim 7$ TeV 
and $46 \lesssim \tan \beta \lesssim 48$, corresponding to a moderate breaking of $t - b$ unification, 
since it corresponds to $0.2 \gtrsim (\la_t/\la_b - 1) \gtrsim 0.1$.

\item The interesting region, characterized in point \ref{itemI}., emerges as a compromise between 
$B \to X_s \gamma$ and $m_b$, pushing $\tan \beta$ to respectively lower and larger values.
In this region, the lightest stop mass is below order 1 TeV, but not less than roughly 800 GeV.

\item If $m_{16}$ is not too large, the interesting region is {\em clearly}
distinguished from the corresponding case with exact Yukawa unification, as far as the fit quality 
is concerned. By looking at panels (a)-(c), one can in fact recognize that the gradient of $\chi^2$ 
variation is close to vertical for, say, $m_{16} = 7$ TeV, when increasing $\tan \beta$ from 
around 48. On the other hand, increasing $m_{16}$ makes the case of breaking $t-b$ unification 
more and more indistinguishable, for the fit performance, from the decoupling 
regime of point 1.

\item \label{itemF}An example of a fit in the interesting region is reported in table \ref{tab:fit-example} 
and shown in panels (a)-(c) as a black square. Note that the prediction for 
BR$(B \to X_s \gamma)$ still tends to be on the lower side of the experimental range from 
eq. (\ref{bsgamma-exp}), with a central value around $2.9 \times 10^{-4}$.
As a comparison, a representative fit in the region with exact Yukawa unification, featuring a flipped 
$b \to s \gamma$ amplitude, is reported in table \ref{tab:flipped-fit-example}.

\end{enumerate}

We note, point 1. above implies that, for any SUSY GUT with Yukawa unification, compatibility with FCNC 
observables and with the now precisely known value of $m_b$ selects the ``partially decoupled 
solution'', advocated in Tobe and Wells \cite{TobeWells}, as the only phenomenologically viable. 
The low-$m_{16}$ solution originally found in \cite{BDR1,BDR2} is disfavoured when combining all 
the most recent data. This conclusion has been here quantitatively assessed with a $\chi^2$ 
procedure.\footnote{Note that, in our approach, exact Yukawa unification can be enforced, so that 
the lower bound on $m_{16}$ emerges transparently as a tension among observables. Instead,
in, e.g., \cite{Auto}, it is low-energy observables (like $m_b$) to be fixed and a large value 
for $m_{16}$ is needed for Yukawa unification to occur within a given tolerance.}

Points \ref{itemI}. to \ref{itemF}. illustrate instead that compatibility among all the 
considered observables at values of $m_{16}$ of order 7 TeV can be recovered at the price of relaxing 
$t - b - \tau$ unification to just $b - \tau$ unification and without modifying universalities in the 
soft terms at the GUT scale.

We conclude this section with a few additional remarks. A first interesting issue is whether 
the general tension involving FCNCs and $m_b$ studied in this Letter may be relieved if, instead or in addition 
to lowering $\tan \beta$, one allows for a complex phase in the trilinear coupling $A_t$. The latter in our 
analysis is real, since we take real $A_0$. A complex phase $\phi_t$ in $A_t$ would induce a $\cos \phi_t$ 
suppression factor in the leading chargino correction to BR$(B \to X_s \gamma)$. In the presence of a complex $A_t$, 
however, the $\tan \be$-enhanced chargino contribution to the bottom mass would also become complex, so that a 
(chiral) redefinition of the down-quark fields would be necessary in order to 
end up with real and positive masses. This may impact non-negligibly the overall size of SUSY corrections to 
$m_b$ \cite{HRS,BDR2} as well as the pattern of CP violation in $B$-physics observables. In addition, a complex 
phase in $A_t$ is also constrained by the electric dipole moments of the electron and neutron. 
Addressing these issues quantitatively goes beyond the scope of the present Letter, where we confine ourselves 
to real GUT-scale soft terms.

As a second remark, we note that, in our analysis of the parameter space, we restricted ourselves to 
positive values of the $\mu$ parameter. This is because, for negative $\mu$, the ISMH solution is 
lost \cite{AABuGuS}, thus leading to much heavier third generation sfermions for a given value of $m_{16}$ 
(typically $m_{\tilde t} \gtrsim 2.5$ TeV), which is unattractive for naturalness reasons. In addition, 
this would lead to a negative contribution to the anomalous magnetic moment of the muon 
$a_\mu$, whereas a sizeable positive contribution is currently favoured by experiment.

In fact, in the region preferred by the $\chi^2$ function, the contributions to $a_\mu$ are 
much smaller than would be needed to explain the E821 result \cite{g-2E821,g-2muon}. This is a 
well-known fact \cite{DR06,AABuGuS} and we decided not to include this observable in our $\chi^2$ 
function to obtain results unbiased by this issue, which still needs to be settled.

We also chose not to include the constraint from the `standard' neutralino relic density 
calculation. The latter yields {\em a posteriori} a too large relic density in the interesting 
region of parameter space for well-known reasons, i.e., the heaviness of sfermions and that of the 
pseudoscalar Higgs, making the $A$-funnel region unattainable. This constraint can however be 
circumvented in many ways, such as a late-time 
entropy injection (see, e.g., \cite{GelminiGondolo}).\footnote{The requirement that SO(10) SUSY GUTs 
with Yukawa unification do account in full for the standard CDM abundance has been recently 
reconsidered in \cite{Baer-CDM}.}

For the above reasons, we preferred to restrict the set of observables considered in the 
analysis to those listed in tables \ref{tab:obs-EW} and \ref{tab:obs-FC}, on which consensus 
is broad.

\section{Conclusions}

In this Letter we have studied the viability of the hypothesis of $t - b - \tau$ Yukawa 
unification in SUSY GUTs under the assumption that soft-breaking terms for sfermions and 
gauginos be universal at the GUT scale. We found that this hypothesis is challenged, unless 
the squark spectrum is pushed well above 1 TeV. Our conclusion is assessed through a global 
fit including EW observables as well as quark FCNC processes. The origin of the difficulty 
is mostly in the specific parameter region chosen by Yukawa unification, which guarantees 
the correct value for the bottom mass and implies ISMH. In this region, it is impossible to 
accommodate simultaneously the experimental value for BR$(B \to X_s \gamma)$ and the severe 
upper bound on BR$(B_s \to \mu^+ \mu^-)$. This statement holds under the prior 
assumption that the sign of the $b \to s \gamma$ amplitude be the same as in the SM. 
For $m_{16} \lesssim 4.7$ TeV, fits prefer the flipped-sign solution for the $b \to s \gamma$ 
amplitude. In this instance it is however difficult to achieve agreement on the $b \to s \gamma$ 
prediction and, simultaneously, on those of EW observables and/or of the bottom mass. In addition, 
this possibility entails in our case a $3\sigma$ discrepancy in $B \to X_s \ell^+ \ell^-$ data.

We have shown that our above conclusions hold irrespectively of the flavour model embedded 
in the SUSY GUT, since, due to ISMH, first and second generation squark masses are much heavier 
than third generation ones.

We have also addressed the possibility of relaxing Yukawa unification to $b - \tau$ 
unification, which in fact allows to maintain most of the predictivity of SUSY GUTs.
In this case, the tension underlined above is in fact largely relieved. The fit still prefers 
large values of $46 \lesssim \tan \beta \lesssim 48$, as a compromise between FCNCs and $m_b$, 
pushing $\tan \beta$ to respectively lower and larger values. The range for $\tan \beta$ 
corresponds to a moderate breaking of $t- b$ Yukawa unification, in the interval
$\la_t/\la_b - 1 \in [0.1,0.2]$. 

In the interesting region, we find the lightest stop mass $\gtrsim 800$ GeV, a light gluino 
around 400 GeV and lightest Higgs, neutralino and chargino close to the lower bounds. This 
spectrum implies BR$(B_s \to \mu^+ \mu^-)$ in the range 2 to 4$\times 10^{-8}$ and 
BR$(B \to X_s \gamma)$, on the lower side of the ``acceptable'' range, $\approx 2.9 \times 10^{-4}$. 
We stress that the requirements of $b -\tau$ unification and the FCNC constraints are enough 
to make the above figures, exemplified in table \ref{tab:fit-example}, basically a firm prediction 
within the interesting region, hence easily falsifiable once the LHC turns on.

\begin{acknowledgments}
We wish to thank A. J. Buras for many useful comments and suggestions. We also acknowledge
M. Ratz for clarifying conversations and remarks. We are indebted to U. Haisch for 
insightful comments, especially useful in the assessment of the $b \to s \gamma$ theoretical error, 
and for providing us with code allowing a number of cross-checks.
The work of W.A., D.G. and D.M.S. has been supported in part by the Cluster of Excellence 
``Origin and Structure of the Universe'' and by the German Bundesministerium f{\"u}r Bildung 
und Forschung under contract 05HT6WOA. D.G. also warmly acknowledges the support of the A. von 
Humboldt Stiftung. S.R. acknowledges partial funding from DOE grant DOE/ER/01545-877.
\end{acknowledgments}

\bibliography{AlGuRaS}

\begin{thebibliography}{62}
\expandafter\ifx\csname natexlab\endcsname\relax\def\natexlab#1{#1}\fi
\expandafter\ifx\csname bibnamefont\endcsname\relax
  \def\bibnamefont#1{#1}\fi
\expandafter\ifx\csname bibfnamefont\endcsname\relax
  \def\bibfnamefont#1{#1}\fi
\expandafter\ifx\csname citenamefont\endcsname\relax
  \def\citenamefont#1{#1}\fi
\expandafter\ifx\csname url\endcsname\relax
  \def\url#1{\texttt{#1}}\fi
\expandafter\ifx\csname urlprefix\endcsname\relax\def\urlprefix{URL }\fi
\providecommand{\bibinfo}[2]{#2}
\providecommand{\eprint}[2][]{\url{#2}}

\bibitem[{\citenamefont{Lucas and Raby}(1997)}]{LucasRabyPRD}
\bibinfo{author}{\bibfnamefont{V.}~\bibnamefont{Lucas}} \bibnamefont{and}
  \bibinfo{author}{\bibfnamefont{S.}~\bibnamefont{Raby}},
  \bibinfo{journal}{Phys. Rev.} \textbf{\bibinfo{volume}{D55}},
  \bibinfo{pages}{6986} (\bibinfo{year}{1997}), \eprint{hep-ph/9610293}.

\bibitem[{\citenamefont{Goto and Nihei}(1999)}]{GotoNihei}
\bibinfo{author}{\bibfnamefont{T.}~\bibnamefont{Goto}} \bibnamefont{and}
  \bibinfo{author}{\bibfnamefont{T.}~\bibnamefont{Nihei}},
  \bibinfo{journal}{Phys. Rev.} \textbf{\bibinfo{volume}{D59}},
  \bibinfo{pages}{115009} (\bibinfo{year}{1999}), \eprint{hep-ph/9808255}.

\bibitem[{\citenamefont{Babu et~al.}(2000)\citenamefont{Babu, Pati, and
  Wilczek}}]{BabuPatiWilczek}
\bibinfo{author}{\bibfnamefont{K.~S.} \bibnamefont{Babu}},
  \bibinfo{author}{\bibfnamefont{J.~C.} \bibnamefont{Pati}}, \bibnamefont{and}
  \bibinfo{author}{\bibfnamefont{F.}~\bibnamefont{Wilczek}},
  \bibinfo{journal}{Nucl. Phys.} \textbf{\bibinfo{volume}{B566}},
  \bibinfo{pages}{33} (\bibinfo{year}{2000}), \eprint{hep-ph/9812538}.

\bibitem[{\citenamefont{Dermisek et~al.}(2001)\citenamefont{Dermisek, Mafi, and
  Raby}}]{DermisekMafiRaby}
\bibinfo{author}{\bibfnamefont{R.}~\bibnamefont{Dermisek}},
  \bibinfo{author}{\bibfnamefont{A.}~\bibnamefont{Mafi}}, \bibnamefont{and}
  \bibinfo{author}{\bibfnamefont{S.}~\bibnamefont{Raby}},
  \bibinfo{journal}{Phys. Rev.} \textbf{\bibinfo{volume}{D63}},
  \bibinfo{pages}{035001} (\bibinfo{year}{2001}), \eprint{hep-ph/0007213}.

\bibitem[{\citenamefont{Dutta et~al.}(2005)\citenamefont{Dutta, Mimura, and
  Mohapatra}}]{Dutta04}
\bibinfo{author}{\bibfnamefont{B.}~\bibnamefont{Dutta}},
  \bibinfo{author}{\bibfnamefont{Y.}~\bibnamefont{Mimura}}, \bibnamefont{and}
  \bibinfo{author}{\bibfnamefont{R.~N.} \bibnamefont{Mohapatra}},
  \bibinfo{journal}{Phys. Rev. Lett.} \textbf{\bibinfo{volume}{94}},
  \bibinfo{pages}{091804} (\bibinfo{year}{2005}), \eprint{hep-ph/0412105}.

\bibitem[{\citenamefont{Dutta et~al.}(2007)\citenamefont{Dutta, Mimura, and
  Mohapatra}}]{Dutta07}
\bibinfo{author}{\bibfnamefont{B.}~\bibnamefont{Dutta}},
  \bibinfo{author}{\bibfnamefont{Y.}~\bibnamefont{Mimura}}, \bibnamefont{and}
  \bibinfo{author}{\bibfnamefont{R.~N.} \bibnamefont{Mohapatra}}
  (\bibinfo{year}{2007}), \eprint{arXiv:0712.1206 [hep-ph]}.

\bibitem[{\citenamefont{Nath and Fileviez~Perez}(2007)}]{Nath06}
\bibinfo{author}{\bibfnamefont{P.}~\bibnamefont{Nath}} \bibnamefont{and}
  \bibinfo{author}{\bibfnamefont{P.}~\bibnamefont{Fileviez~Perez}},
  \bibinfo{journal}{Phys. Rept.} \textbf{\bibinfo{volume}{441}},
  \bibinfo{pages}{191} (\bibinfo{year}{2007}), \eprint{hep-ph/0601023}.

\bibitem[{\citenamefont{Nath and Syed}(2007)}]{Nath07}
\bibinfo{author}{\bibfnamefont{P.}~\bibnamefont{Nath}} \bibnamefont{and}
  \bibinfo{author}{\bibfnamefont{R.~M.} \bibnamefont{Syed}}
  (\bibinfo{year}{2007}), \eprint{arXiv:0707.1332 [hep-ph]}.

\bibitem[{\citenamefont{Albrecht et~al.}(2007)\citenamefont{Albrecht,
  Altmannshofer, Buras, Guadagnoli, and Straub}}]{AABuGuS}
\bibinfo{author}{\bibfnamefont{M.~E.} \bibnamefont{Albrecht}},
  \bibinfo{author}{\bibfnamefont{W.}~\bibnamefont{Altmannshofer}},
  \bibinfo{author}{\bibfnamefont{A.~J.} \bibnamefont{Buras}},
  \bibinfo{author}{\bibfnamefont{D.}~\bibnamefont{Guadagnoli}},
  \bibnamefont{and} \bibinfo{author}{\bibfnamefont{D.~M.}
  \bibnamefont{Straub}}, \bibinfo{journal}{JHEP} \textbf{\bibinfo{volume}{10}},
  \bibinfo{pages}{055} (\bibinfo{year}{2007}), \eprint{arXiv:0707.3954
  [hep-ph]}.

\bibitem[{\citenamefont{Dermisek and Raby}(2005)}]{DR05}
\bibinfo{author}{\bibfnamefont{R.}~\bibnamefont{Dermisek}} \bibnamefont{and}
  \bibinfo{author}{\bibfnamefont{S.}~\bibnamefont{Raby}},
  \bibinfo{journal}{Phys. Lett.} \textbf{\bibinfo{volume}{B622}},
  \bibinfo{pages}{327} (\bibinfo{year}{2005}), \eprint{hep-ph/0507045}.

\bibitem[{\citenamefont{Dermisek et~al.}(2006)\citenamefont{Dermisek, Harada,
  and Raby}}]{DR06}
\bibinfo{author}{\bibfnamefont{R.}~\bibnamefont{Dermisek}},
  \bibinfo{author}{\bibfnamefont{M.}~\bibnamefont{Harada}}, \bibnamefont{and}
  \bibinfo{author}{\bibfnamefont{S.}~\bibnamefont{Raby}},
  \bibinfo{journal}{Phys. Rev.} \textbf{\bibinfo{volume}{D74}},
  \bibinfo{pages}{035011} (\bibinfo{year}{2006}), \eprint{hep-ph/0606055}.

\bibitem[{\citenamefont{Blazek et~al.}(2002{\natexlab{a}})\citenamefont{Blazek,
  Dermisek, and Raby}}]{BDR1}
\bibinfo{author}{\bibfnamefont{T.}~\bibnamefont{Blazek}},
  \bibinfo{author}{\bibfnamefont{R.}~\bibnamefont{Dermisek}}, \bibnamefont{and}
  \bibinfo{author}{\bibfnamefont{S.}~\bibnamefont{Raby}},
  \bibinfo{journal}{Phys. Rev. Lett.} \textbf{\bibinfo{volume}{88}},
  \bibinfo{pages}{111804} (\bibinfo{year}{2002}{\natexlab{a}}),
  \eprint{hep-ph/0107097}.

\bibitem[{\citenamefont{Blazek et~al.}(2002{\natexlab{b}})\citenamefont{Blazek,
  Dermisek, and Raby}}]{BDR2}
\bibinfo{author}{\bibfnamefont{T.}~\bibnamefont{Blazek}},
  \bibinfo{author}{\bibfnamefont{R.}~\bibnamefont{Dermisek}}, \bibnamefont{and}
  \bibinfo{author}{\bibfnamefont{S.}~\bibnamefont{Raby}},
  \bibinfo{journal}{Phys. Rev.} \textbf{\bibinfo{volume}{D65}},
  \bibinfo{pages}{115004} (\bibinfo{year}{2002}{\natexlab{b}}),
  \eprint{hep-ph/0201081}.

\bibitem[{\citenamefont{Hall et~al.}(1994)\citenamefont{Hall, Rattazzi, and
  Sarid}}]{HRS}
\bibinfo{author}{\bibfnamefont{L.~J.} \bibnamefont{Hall}},
  \bibinfo{author}{\bibfnamefont{R.}~\bibnamefont{Rattazzi}}, \bibnamefont{and}
  \bibinfo{author}{\bibfnamefont{U.}~\bibnamefont{Sarid}},
  \bibinfo{journal}{Phys. Rev.} \textbf{\bibinfo{volume}{D50}},
  \bibinfo{pages}{7048} (\bibinfo{year}{1994}), \eprint{hep-ph/9306309}.

\bibitem[{\citenamefont{Bagger et~al.}(2000)\citenamefont{Bagger, Feng,
  Polonsky, and Zhang}}]{BFPZ}
\bibinfo{author}{\bibfnamefont{J.~A.} \bibnamefont{Bagger}},
  \bibinfo{author}{\bibfnamefont{J.~L.} \bibnamefont{Feng}},
  \bibinfo{author}{\bibfnamefont{N.}~\bibnamefont{Polonsky}}, \bibnamefont{and}
  \bibinfo{author}{\bibfnamefont{R.-J.} \bibnamefont{Zhang}},
  \bibinfo{journal}{Phys. Lett.} \textbf{\bibinfo{volume}{B473}},
  \bibinfo{pages}{264} (\bibinfo{year}{2000}), \eprint{hep-ph/9911255}.

\bibitem[{\citenamefont{Babu and Kolda}(2000)}]{BabuKolda}
\bibinfo{author}{\bibfnamefont{K.~S.} \bibnamefont{Babu}} \bibnamefont{and}
  \bibinfo{author}{\bibfnamefont{C.~F.} \bibnamefont{Kolda}},
  \bibinfo{journal}{Phys. Rev. Lett.} \textbf{\bibinfo{volume}{84}},
  \bibinfo{pages}{228} (\bibinfo{year}{2000}), \eprint{hep-ph/9909476}.

\bibitem[{\citenamefont{Choudhury and Gaur}(1999)}]{Gaur}
\bibinfo{author}{\bibfnamefont{S.~R.} \bibnamefont{Choudhury}}
  \bibnamefont{and} \bibinfo{author}{\bibfnamefont{N.}~\bibnamefont{Gaur}},
  \bibinfo{journal}{Phys. Lett.} \textbf{\bibinfo{volume}{B451}},
  \bibinfo{pages}{86} (\bibinfo{year}{1999}), \eprint{hep-ph/9810307}.

\bibitem[{\citenamefont{Aaltonen et~al.}(2007)}]{Bsmumu-CDF}
\bibinfo{author}{\bibfnamefont{T.}~\bibnamefont{Aaltonen}} \bibnamefont{et~al.}
  (\bibinfo{collaboration}{CDF}) (\bibinfo{year}{2007}),
  \eprint{arXiv:0712.1708 [hep-ex]}.

\bibitem[{UTf()}]{UTfit}
\bibinfo{note}{{UT$_{fit}$ website: {\sf www.utfit.org}}}.

\bibitem[{CKM()}]{CKMfitter}
\bibinfo{note}{{CKMfitter website: {\sf ckmfitter.in2p3.fr}}}.

\bibitem[{Ika()}]{Ikado}
\bibinfo{note}{K. Ikado, talk presented at FPCP 2006 (9-12 April 2006,
  Vancouver, Canada) {\sf fpcp2006.triumf.ca}.}

\bibitem[{\citenamefont{Carena et~al.}(2007)\citenamefont{Carena, Menon, and
  Wagner}}]{CMW}
\bibinfo{author}{\bibfnamefont{M.~S.} \bibnamefont{Carena}},
  \bibinfo{author}{\bibfnamefont{A.}~\bibnamefont{Menon}}, \bibnamefont{and}
  \bibinfo{author}{\bibfnamefont{C.~E.~M.} \bibnamefont{Wagner}},
  \bibinfo{journal}{Phys. Rev.} \textbf{\bibinfo{volume}{D76}},
  \bibinfo{pages}{035004} (\bibinfo{year}{2007}), \eprint{arXiv:0704.1143
  [hep-ph]}.

\bibitem[{HFA()}]{HFAG}
\bibinfo{note}{{Heavy Flavor Averaging Group:\\ {\sf
  www.slac.stanford.edu/xorg/hfag}}}.

\bibitem[{\citenamefont{Aubert et~al.}(2007)}]{Babar-Btaunu}
\bibinfo{author}{\bibfnamefont{B.}~\bibnamefont{Aubert}} \bibnamefont{et~al.}
  (\bibinfo{collaboration}{BABAR}), \bibinfo{journal}{Phys. Rev.}
  \textbf{\bibinfo{volume}{D76}}, \bibinfo{pages}{052002}
  (\bibinfo{year}{2007}), \eprint{arXiv:0708.2260 [hep-ex]}.

\bibitem[{\citenamefont{Ikado et~al.}(2006)}]{Belle-Btaunu}
\bibinfo{author}{\bibfnamefont{K.}~\bibnamefont{Ikado}} \bibnamefont{et~al.},
  \bibinfo{journal}{Phys. Rev. Lett.} \textbf{\bibinfo{volume}{97}},
  \bibinfo{pages}{251802} (\bibinfo{year}{2006}), \eprint{hep-ex/0604018}.

\bibitem[{\citenamefont{Hou}(1993)}]{Hou}
\bibinfo{author}{\bibfnamefont{W.-S.} \bibnamefont{Hou}},
  \bibinfo{journal}{Phys. Rev.} \textbf{\bibinfo{volume}{D48}},
  \bibinfo{pages}{2342} (\bibinfo{year}{1993}).

\bibitem[{\citenamefont{Barberio et~al.}(2007)}]{Barberio07}
\bibinfo{author}{\bibfnamefont{E.}~\bibnamefont{Barberio}} \bibnamefont{et~al.}
  (\bibinfo{collaboration}{Heavy Flavor Averaging Group (HFAG)})
  (\bibinfo{year}{2007}), \eprint{arXiv:0704.3575 [hep-ex]}.

\bibitem[{\citenamefont{Misiak et~al.}(2007)}]{Misiak-NNLO}
\bibinfo{author}{\bibfnamefont{M.}~\bibnamefont{Misiak}} \bibnamefont{et~al.},
  \bibinfo{journal}{Phys. Rev. Lett.} \textbf{\bibinfo{volume}{98}},
  \bibinfo{pages}{022002} (\bibinfo{year}{2007}), \eprint{hep-ph/0609232}.

\bibitem[{\citenamefont{Blazek and Raby}(1999)}]{BlazekRaby-bsgamma}
\bibinfo{author}{\bibfnamefont{T.}~\bibnamefont{Blazek}} \bibnamefont{and}
  \bibinfo{author}{\bibfnamefont{S.}~\bibnamefont{Raby}},
  \bibinfo{journal}{Phys. Rev.} \textbf{\bibinfo{volume}{D59}},
  \bibinfo{pages}{095002} (\bibinfo{year}{1999}), \eprint{hep-ph/9712257}.

\bibitem[{\citenamefont{Gambino et~al.}(2005)\citenamefont{Gambino, Haisch, and
  Misiak}}]{GHM}
\bibinfo{author}{\bibfnamefont{P.}~\bibnamefont{Gambino}},
  \bibinfo{author}{\bibfnamefont{U.}~\bibnamefont{Haisch}}, \bibnamefont{and}
  \bibinfo{author}{\bibfnamefont{M.}~\bibnamefont{Misiak}},
  \bibinfo{journal}{Phys. Rev. Lett.} \textbf{\bibinfo{volume}{94}},
  \bibinfo{pages}{061803} (\bibinfo{year}{2005}), \eprint{hep-ph/0410155}.

\bibitem[{\citenamefont{Ali et~al.}(2002)\citenamefont{Ali, Lunghi, Greub, and
  Hiller}}]{ALGH}
\bibinfo{author}{\bibfnamefont{A.}~\bibnamefont{Ali}},
  \bibinfo{author}{\bibfnamefont{E.}~\bibnamefont{Lunghi}},
  \bibinfo{author}{\bibfnamefont{C.}~\bibnamefont{Greub}}, \bibnamefont{and}
  \bibinfo{author}{\bibfnamefont{G.}~\bibnamefont{Hiller}},
  \bibinfo{journal}{Phys. Rev.} \textbf{\bibinfo{volume}{D66}},
  \bibinfo{pages}{034002} (\bibinfo{year}{2002}), \eprint{hep-ph/0112300}.

\bibitem[{\citenamefont{Lunghi et~al.}(2006)\citenamefont{Lunghi, Porod, and
  Vives}}]{LPV}
\bibinfo{author}{\bibfnamefont{E.}~\bibnamefont{Lunghi}},
  \bibinfo{author}{\bibfnamefont{W.}~\bibnamefont{Porod}}, \bibnamefont{and}
  \bibinfo{author}{\bibfnamefont{O.}~\bibnamefont{Vives}},
  \bibinfo{journal}{Phys. Rev.} \textbf{\bibinfo{volume}{D74}},
  \bibinfo{pages}{075003} (\bibinfo{year}{2006}), \eprint{hep-ph/0605177}.

\bibitem[{\citenamefont{Barr and Raby}(1997)}]{BarrRaby}
\bibinfo{author}{\bibfnamefont{S.~M.} \bibnamefont{Barr}} \bibnamefont{and}
  \bibinfo{author}{\bibfnamefont{S.}~\bibnamefont{Raby}},
  \bibinfo{journal}{Phys. Rev. Lett.} \textbf{\bibinfo{volume}{79}},
  \bibinfo{pages}{4748} (\bibinfo{year}{1997}), \eprint{hep-ph/9705366}.

\bibitem[{\citenamefont{Albright and Barr}(2000)}]{AlbrightBarr}
\bibinfo{author}{\bibfnamefont{C.~H.} \bibnamefont{Albright}} \bibnamefont{and}
  \bibinfo{author}{\bibfnamefont{S.~M.} \bibnamefont{Barr}},
  \bibinfo{journal}{Phys. Rev.} \textbf{\bibinfo{volume}{D62}},
  \bibinfo{pages}{093008} (\bibinfo{year}{2000}), \eprint{hep-ph/0003251}.

\bibitem[{\citenamefont{Schael et~al.}(2006)}]{LEPHiggs}
\bibinfo{author}{\bibfnamefont{S.}~\bibnamefont{Schael}} \bibnamefont{et~al.}
  (\bibinfo{collaboration}{LEP Higgs Working Group}), \bibinfo{journal}{Eur.
  Phys. J.} \textbf{\bibinfo{volume}{C47}}, \bibinfo{pages}{547}
  (\bibinfo{year}{2006}), \eprint{hep-ex/0602042}.

\bibitem[{\citenamefont{Carena et~al.}(1993)\citenamefont{Carena, Pokorski, and
  Wagner}}]{CPW}
\bibinfo{author}{\bibfnamefont{M.~S.} \bibnamefont{Carena}},
  \bibinfo{author}{\bibfnamefont{S.}~\bibnamefont{Pokorski}}, \bibnamefont{and}
  \bibinfo{author}{\bibfnamefont{C.~E.~M.} \bibnamefont{Wagner}},
  \bibinfo{journal}{Nucl. Phys.} \textbf{\bibinfo{volume}{B406}},
  \bibinfo{pages}{59} (\bibinfo{year}{1993}), \eprint{hep-ph/9303202}.

\bibitem[{\citenamefont{Ananthanarayan
  et~al.}(1994)\citenamefont{Ananthanarayan, Babu, and Shafi}}]{ananthanarayan}
\bibinfo{author}{\bibfnamefont{B.}~\bibnamefont{Ananthanarayan}},
  \bibinfo{author}{\bibfnamefont{K.~S.} \bibnamefont{Babu}}, \bibnamefont{and}
  \bibinfo{author}{\bibfnamefont{Q.}~\bibnamefont{Shafi}},
  \bibinfo{journal}{Nucl. Phys.} \textbf{\bibinfo{volume}{B428}},
  \bibinfo{pages}{19} (\bibinfo{year}{1994}), \eprint{hep-ph/9402284}.

\bibitem[{\citenamefont{Martin and Vaughn}(1994)}]{MartinVaughn}
\bibinfo{author}{\bibfnamefont{S.~P.} \bibnamefont{Martin}} \bibnamefont{and}
  \bibinfo{author}{\bibfnamefont{M.~T.} \bibnamefont{Vaughn}},
  \bibinfo{journal}{Phys. Rev.} \textbf{\bibinfo{volume}{D50}},
  \bibinfo{pages}{2282} (\bibinfo{year}{1994}), \eprint{hep-ph/9311340}.

\bibitem[{\citenamefont{Hisano et~al.}(1996)\citenamefont{Hisano, Moroi, Tobe,
  and Yamaguchi}}]{Hisano95}
\bibinfo{author}{\bibfnamefont{J.}~\bibnamefont{Hisano}},
  \bibinfo{author}{\bibfnamefont{T.}~\bibnamefont{Moroi}},
  \bibinfo{author}{\bibfnamefont{K.}~\bibnamefont{Tobe}}, \bibnamefont{and}
  \bibinfo{author}{\bibfnamefont{M.}~\bibnamefont{Yamaguchi}},
  \bibinfo{journal}{Phys. Rev.} \textbf{\bibinfo{volume}{D53}},
  \bibinfo{pages}{2442} (\bibinfo{year}{1996}), \eprint{hep-ph/9510309}.

\bibitem[{\citenamefont{Antusch et~al.}(2002)\citenamefont{Antusch, Kersten,
  Lindner, and Ratz}}]{AKLR02}
\bibinfo{author}{\bibfnamefont{S.}~\bibnamefont{Antusch}},
  \bibinfo{author}{\bibfnamefont{J.}~\bibnamefont{Kersten}},
  \bibinfo{author}{\bibfnamefont{M.}~\bibnamefont{Lindner}}, \bibnamefont{and}
  \bibinfo{author}{\bibfnamefont{M.}~\bibnamefont{Ratz}},
  \bibinfo{journal}{Phys. Lett.} \textbf{\bibinfo{volume}{B538}},
  \bibinfo{pages}{87} (\bibinfo{year}{2002}), \eprint{hep-ph/0203233}.

\bibitem[{\citenamefont{Petcov et~al.}(2004)\citenamefont{Petcov, Profumo,
  Takanishi, and Yaguna}}]{Petcov03}
\bibinfo{author}{\bibfnamefont{S.~T.} \bibnamefont{Petcov}},
  \bibinfo{author}{\bibfnamefont{S.}~\bibnamefont{Profumo}},
  \bibinfo{author}{\bibfnamefont{Y.}~\bibnamefont{Takanishi}},
  \bibnamefont{and} \bibinfo{author}{\bibfnamefont{C.~E.}
  \bibnamefont{Yaguna}}, \bibinfo{journal}{Nucl. Phys.}
  \textbf{\bibinfo{volume}{B676}}, \bibinfo{pages}{453} (\bibinfo{year}{2004}),
  \eprint{hep-ph/0306195}.

\bibitem[{\citenamefont{Pierce et~al.}(1997)\citenamefont{Pierce, Bagger,
  Matchev, and Zhang}}]{PBMZ}
\bibinfo{author}{\bibfnamefont{D.~M.} \bibnamefont{Pierce}},
  \bibinfo{author}{\bibfnamefont{J.~A.} \bibnamefont{Bagger}},
  \bibinfo{author}{\bibfnamefont{K.~T.} \bibnamefont{Matchev}},
  \bibnamefont{and} \bibinfo{author}{\bibfnamefont{R.-j.} \bibnamefont{Zhang}},
  \bibinfo{journal}{Nucl. Phys.} \textbf{\bibinfo{volume}{B491}},
  \bibinfo{pages}{3} (\bibinfo{year}{1997}), \eprint{hep-ph/9606211}.

\bibitem[{\citenamefont{Heinemeyer et~al.}(2000)\citenamefont{Heinemeyer,
  Hollik, and Weiglein}}]{FeynHiggs1}
\bibinfo{author}{\bibfnamefont{S.}~\bibnamefont{Heinemeyer}},
  \bibinfo{author}{\bibfnamefont{W.}~\bibnamefont{Hollik}}, \bibnamefont{and}
  \bibinfo{author}{\bibfnamefont{G.}~\bibnamefont{Weiglein}},
  \bibinfo{journal}{Comput. Phys. Commun.} \textbf{\bibinfo{volume}{124}},
  \bibinfo{pages}{76} (\bibinfo{year}{2000}), \eprint{hep-ph/9812320}.

\bibitem[{\citenamefont{Heinemeyer et~al.}(1999)\citenamefont{Heinemeyer,
  Hollik, and Weiglein}}]{FeynHiggs2}
\bibinfo{author}{\bibfnamefont{S.}~\bibnamefont{Heinemeyer}},
  \bibinfo{author}{\bibfnamefont{W.}~\bibnamefont{Hollik}}, \bibnamefont{and}
  \bibinfo{author}{\bibfnamefont{G.}~\bibnamefont{Weiglein}},
  \bibinfo{journal}{Eur. Phys. J.} \textbf{\bibinfo{volume}{C9}},
  \bibinfo{pages}{343} (\bibinfo{year}{1999}), \eprint{hep-ph/9812472}.

\bibitem[{\citenamefont{Degrassi et~al.}(2003)\citenamefont{Degrassi,
  Heinemeyer, Hollik, Slavich, and Weiglein}}]{FeynHiggs3}
\bibinfo{author}{\bibfnamefont{G.}~\bibnamefont{Degrassi}},
  \bibinfo{author}{\bibfnamefont{S.}~\bibnamefont{Heinemeyer}},
  \bibinfo{author}{\bibfnamefont{W.}~\bibnamefont{Hollik}},
  \bibinfo{author}{\bibfnamefont{P.}~\bibnamefont{Slavich}}, \bibnamefont{and}
  \bibinfo{author}{\bibfnamefont{G.}~\bibnamefont{Weiglein}},
  \bibinfo{journal}{Eur. Phys. J.} \textbf{\bibinfo{volume}{C28}},
  \bibinfo{pages}{133} (\bibinfo{year}{2003}), \eprint{hep-ph/0212020}.

\bibitem[{\citenamefont{Frank et~al.}(2006)}]{FeynHiggs4}
\bibinfo{author}{\bibfnamefont{M.}~\bibnamefont{Frank}} \bibnamefont{et~al.}
  (\bibinfo{year}{2006}), \eprint{hep-ph/0611326}.

\bibitem[{\citenamefont{Buras et~al.}(2003)\citenamefont{Buras, Chankowski,
  Rosiek, and Slawianowska}}]{BCRSbig}
\bibinfo{author}{\bibfnamefont{A.~J.} \bibnamefont{Buras}},
  \bibinfo{author}{\bibfnamefont{P.~H.} \bibnamefont{Chankowski}},
  \bibinfo{author}{\bibfnamefont{J.}~\bibnamefont{Rosiek}}, \bibnamefont{and}
  \bibinfo{author}{\bibfnamefont{L.}~\bibnamefont{Slawianowska}},
  \bibinfo{journal}{Nucl. Phys.} \textbf{\bibinfo{volume}{B659}},
  \bibinfo{pages}{3} (\bibinfo{year}{2003}), \eprint{hep-ph/0210145}.

\bibitem[{\citenamefont{{Yao} et~al.}(2006)}]{PDBook}
\bibinfo{author}{\bibfnamefont{W.-M.} \bibnamefont{{Yao}}}
  \bibnamefont{et~al.}, \bibinfo{journal}{{Journal of Physics G}}
  \textbf{\bibinfo{volume}{33}}, \bibinfo{pages}{{1+}} (\bibinfo{year}{2006}),
  \urlprefix\url{pdg.lbl.gov}.

\bibitem[{\citenamefont{{Tevatron Electroweak Working Group}}(2007)}]{CDFD0mt}
\bibinfo{author}{\bibnamefont{{Tevatron Electroweak Working Group}}}
  (\bibinfo{year}{2007}), \eprint{hep-ex/0703034}.

\bibitem[{\citenamefont{Abulencia et~al.}(2006)}]{CDF-DMs}
\bibinfo{author}{\bibfnamefont{A.}~\bibnamefont{Abulencia}}
  \bibnamefont{et~al.} (\bibinfo{collaboration}{CDF}), \bibinfo{journal}{Phys.
  Rev. Lett.} \textbf{\bibinfo{volume}{97}}, \bibinfo{pages}{242003}
  (\bibinfo{year}{2006}), \eprint{hep-ex/0609040}.

\bibitem[{\citenamefont{Aubert et~al.}(2004)}]{Babar-bsll}
\bibinfo{author}{\bibfnamefont{B.}~\bibnamefont{Aubert}} \bibnamefont{et~al.}
  (\bibinfo{collaboration}{BABAR}), \bibinfo{journal}{Phys. Rev. Lett.}
  \textbf{\bibinfo{volume}{93}}, \bibinfo{pages}{081802}
  (\bibinfo{year}{2004}), \eprint{hep-ex/0404006}.

\bibitem[{\citenamefont{Iwasaki et~al.}(2005)}]{Belle-bsll}
\bibinfo{author}{\bibfnamefont{M.}~\bibnamefont{Iwasaki}} \bibnamefont{et~al.}
  (\bibinfo{collaboration}{Belle}), \bibinfo{journal}{Phys. Rev.}
  \textbf{\bibinfo{volume}{D72}}, \bibinfo{pages}{092005}
  (\bibinfo{year}{2005}), \eprint{hep-ex/0503044}.

\bibitem[{CER()}]{CERNlib}
\bibinfo{note}{{See the {\tt CERNlib} website: {\sf
  cernlib.web.cern.ch/cernlib/}}}.

\bibitem[{\citenamefont{Degrassi et~al.}(2007)\citenamefont{Degrassi, Gambino,
  and Slavich}}]{SusyBSG}
\bibinfo{author}{\bibfnamefont{G.}~\bibnamefont{Degrassi}},
  \bibinfo{author}{\bibfnamefont{P.}~\bibnamefont{Gambino}}, \bibnamefont{and}
  \bibinfo{author}{\bibfnamefont{P.}~\bibnamefont{Slavich}}
  (\bibinfo{year}{2007}), \eprint{arXiv:0712.3265 [hep-ph]}.

\bibitem[{\citenamefont{Huber et~al.}(2006)\citenamefont{Huber, Lunghi, Misiak,
  and Wyler}}]{Huber-et-al}
\bibinfo{author}{\bibfnamefont{T.}~\bibnamefont{Huber}},
  \bibinfo{author}{\bibfnamefont{E.}~\bibnamefont{Lunghi}},
  \bibinfo{author}{\bibfnamefont{M.}~\bibnamefont{Misiak}}, \bibnamefont{and}
  \bibinfo{author}{\bibfnamefont{D.}~\bibnamefont{Wyler}},
  \bibinfo{journal}{Nucl. Phys.} \textbf{\bibinfo{volume}{B740}},
  \bibinfo{pages}{105} (\bibinfo{year}{2006}), \eprint{hep-ph/0512066}.

\bibitem[{\citenamefont{Roszkowski et~al.}(2007)\citenamefont{Roszkowski,
  Ruiz~de Austri, and Trotta}}]{Roszkowski-CMSSM}
\bibinfo{author}{\bibfnamefont{L.}~\bibnamefont{Roszkowski}},
  \bibinfo{author}{\bibfnamefont{R.}~\bibnamefont{Ruiz~de Austri}},
  \bibnamefont{and} \bibinfo{author}{\bibfnamefont{R.}~\bibnamefont{Trotta}},
  \bibinfo{journal}{JHEP} \textbf{\bibinfo{volume}{07}}, \bibinfo{pages}{075}
  (\bibinfo{year}{2007}), \eprint{arXiv:0705.2012 [hep-ph]}.

\bibitem[{\citenamefont{Tobe and Wells}(2003)}]{TobeWells}
\bibinfo{author}{\bibfnamefont{K.}~\bibnamefont{Tobe}} \bibnamefont{and}
  \bibinfo{author}{\bibfnamefont{J.~D.} \bibnamefont{Wells}},
  \bibinfo{journal}{Nucl. Phys.} \textbf{\bibinfo{volume}{B663}},
  \bibinfo{pages}{123} (\bibinfo{year}{2003}), \eprint{hep-ph/0301015}.

\bibitem[{\citenamefont{Auto et~al.}(2003)}]{Auto}
\bibinfo{author}{\bibfnamefont{D.}~\bibnamefont{Auto}} \bibnamefont{et~al.},
  \bibinfo{journal}{JHEP} \textbf{\bibinfo{volume}{06}}, \bibinfo{pages}{023}
  (\bibinfo{year}{2003}), \eprint{hep-ph/0302155}.

\bibitem[{\citenamefont{Bennett et~al.}(2006)}]{g-2E821}
\bibinfo{author}{\bibfnamefont{G.~W.} \bibnamefont{Bennett}}
  \bibnamefont{et~al.} (\bibinfo{collaboration}{Muon ($g-2$)}),
  \bibinfo{journal}{Phys. Rev.} \textbf{\bibinfo{volume}{D73}},
  \bibinfo{pages}{072003} (\bibinfo{year}{2006}), \eprint{hep-ex/0602035}.

\bibitem[{\citenamefont{Miller et~al.}(2007)\citenamefont{Miller, de~Rafael,
  and Roberts}}]{g-2muon}
\bibinfo{author}{\bibfnamefont{J.~P.} \bibnamefont{Miller}},
  \bibinfo{author}{\bibfnamefont{E.}~\bibnamefont{de~Rafael}},
  \bibnamefont{and} \bibinfo{author}{\bibfnamefont{B.~L.}
  \bibnamefont{Roberts}}, \bibinfo{journal}{Rept. Prog. Phys.}
  \textbf{\bibinfo{volume}{70}}, \bibinfo{pages}{795} (\bibinfo{year}{2007}),
  \eprint{hep-ph/0703049}.

\bibitem[{\citenamefont{Gelmini and Gondolo}(2006)}]{GelminiGondolo}
\bibinfo{author}{\bibfnamefont{G.~B.} \bibnamefont{Gelmini}} \bibnamefont{and}
  \bibinfo{author}{\bibfnamefont{P.}~\bibnamefont{Gondolo}},
  \bibinfo{journal}{Phys. Rev.} \textbf{\bibinfo{volume}{D74}},
  \bibinfo{pages}{023510} (\bibinfo{year}{2006}), \eprint{hep-ph/0602230}.

\bibitem[{\citenamefont{Baer et~al.}(2008)\citenamefont{Baer, Kraml, Sekmen,
  and Summy}}]{Baer-CDM}
\bibinfo{author}{\bibfnamefont{H.}~\bibnamefont{Baer}},
  \bibinfo{author}{\bibfnamefont{S.}~\bibnamefont{Kraml}},
  \bibinfo{author}{\bibfnamefont{S.}~\bibnamefont{Sekmen}}, \bibnamefont{and}
  \bibinfo{author}{\bibfnamefont{H.}~\bibnamefont{Summy}}
  (\bibinfo{year}{2008}), \eprint{arXiv:0801.1831 [Unknown]}.

\end{thebibliography}

\end{document}